\documentclass[12pt,cite,a4paper]{article}
\usepackage{myart}
\usepackage{amsmath}
\usepackage{graphicx}
\usepackage{amsfonts}
\usepackage{amssymb}

\begin{document}
\hfill\hbox{RUNHETC-2001-02}

\hfill\hbox{LPM-01-01}

\hfill\hbox{January, 2001}

\bigskip

\begin{center}
{\Large \textbf{Liouville Field Theory on a Pseudosphere}}

\vspace{1.5cm}

{\large A.Zamolodchikov}

\vspace{0.2cm}

and

\vspace{0.2cm}

{\large Al.Zamolodchikov}\footnote{On leave of absence from: Laboratoire de
Physique Math\'ematique (Laboratoire Associ\'e au CNRS URA-768), Universit\'e
Montpellier II, Pl.E.Bataillon, 34095 Montpellier, France and Institute of
Theoretical and Experimental Physics, B.Cheremushkinskaya 25, 117259 Moscow, Russia}

\vspace{0.3cm}

Department of Physics and Astronomy

Rutgers University

P.O.Box 849, Piscataway, New Jersey 08855-0849, USA
\end{center}

\vspace{1.0cm}

\textbf{Abstract}

Liouville field theory is considered with boundary conditions corresponding to
a quantization of the classical Lobachevskiy plane (i.e. euclidean version of
$AdS_2$). We solve the bootstrap
equations for the out-vacuum wave function and find an infinite set of
solutions. This solutions are in one to one correspondence with the degenerate
representations of the Virasoro algebra. Consistency of these solutions is
verified by both boundary and modular bootstrap techniques. Perturbative
calculations lead to the conclusion that only the ``basic'' solution
corresponding to the identity operator provides a ``natural'' quantization of
the Lobachevskiy plane.

\section{Introduction}

Liouville field theory (LFT) is widely considered as an appropriate field
theoretic background for a certain universality class of two-dimensional
quantum gravity. It has been demonstrated in numerous examples that in 2D the
scaling limit of the so-called ``dynamical triangulations''
\cite{Kazakov,Kazakov2,David85,Frohlich} (which are in fact a discrete model
of a two-dimensional surface with fluctuating geometry) in many cases can be
described by appropriately applied LFT \cite{KPZ,David,Distler}

Local dynamics of LFT is determined by the action density
\begin{equation}
\mathcal{L}(z)=\frac1{4\pi}(\partial_{a}\phi(z))^{2}+\mu e^{2b\phi
(z)}\label{LFT}%
\end{equation}
where $\phi$ is the Liouville field and $b$ is a dimensionless parameter
which, roughly speaking, determines the ``rigidity'' of a 2D surface to
quantum fluctuations of the metric. Ordinarily $\exp(2b\phi(z))d^{2}z$ is
interpreted as the quantum volume element of the fluctuating surface,
parameter $\mu$ being the cosmological coupling constant. LFT is a conformal
field theory with central charge
\begin{equation}
c_{L}=1+6Q^{2}\label{cL}%
\end{equation}
where $Q$ is yet another convenient parameter called the ``background
charge''
\begin{equation}
Q=b^{-1}+b\label{Q}%
\end{equation}
More details about the space of states in LFT, the set of local primary fields
and local operator algebra can be found e.g. in \cite{AAl}.

Local equation of motion for (\ref{LFT})
\begin{equation}
\Delta\phi=4\pi\mu be^{2b\phi}\label{Liouv}%
\end{equation}
is the quantum version of the classic Liouville equation
\begin{equation}
\Delta\varphi=2R^{-2}e^{\varphi}\label{liouv}%
\end{equation}
It describes locally a metric
\begin{equation}
ds^{2}=e^{\varphi(z)}\left|  dz\right|  ^{2}\label{ds}%
\end{equation}
of constant negative curvature $-2R^{-2}$ in the isothermal (or conformal)
coordinates. Classical situation arises in LFT\ if $b\rightarrow0$. In this
limit we identify the classical field $\varphi=2b\phi$ while $R^{-2}=4\pi\mu
b^{2}$.

Discrete models of quantum gravity, such as the random triangulations or
matrix models, typically deal with compact fluctuating surfaces of different
topologies either with or without boundaries. This type of problems is most
relevant in the string theory. In this context the main problem is somewhat
different from that considered usually in field theory. Namely, observables of
primary interest are the ``integrated'' correlation functions, which bear no
coordinate dependence and can be rather called the ``correlation numbers''.
They are used to describe certain ``deformations'' or ``flows'' caused by
relevant perturbations (see e.g. the reviews \cite{Klebanov, Ginsparg} and
\cite{DiFrancesco} for more details). In many problems of this kind the
discrete approaches presently appear more efficient then the field theoretic
description based on LFT. In the discrete schemes the correlation functions
naturally arise in the ``integrated'' form while the field theoretic approach
implies a gauge fixing and gives the correlation functions as the functions of
certain moduli (invariants of the complex structure in the case of LFT). These
functions, although being themselves of considerable interest, should be yet
integrated over the moduli space to produce the correlation numbers. Therefore
in such problems of quantum gravity LFT still lags behind matrix models or
other discrete approaches. Up to now only general scaling exponents and a
limited set of correlation functions in simplest compact topologies can be
predicted in LFT (see e.g. \cite{Ginsparg} and references therein).

It is well known that the Liouville equation (\ref{liouv}) admits ``basic''
solution, which describes the geometry of infinite constant negative curvature
surface, the so-called Lobachevskiy plane, or pseudosphere. This surface can be 
realized as the disk $\left|  z\right|  <1$ with metric (\ref{ds}) where
\begin{equation}
e^{\varphi(z)}=\frac{4R^{2}}{(1-z\bar z)^{2}}\label{psphere}%
\end{equation}
Here $R$ is interpreted as the radius of the pseudosphere. The points at the
circle $\left|  z\right|  =1$ are infinitely far away from any internal point
and form a one-dimensional infinity called the absolute. Geometry of the
pseudosphere is described in detail in standard textbooks and here we will not
go into further details like $SL(2,R)$ symmetry, geodesics etc. Let us mention
only the so-called Poincar\'e model, where the same geometry is represented in
the upper half plane of complex $\xi$ with the metric
\begin{equation}
e^{\varphi(\xi)}=\frac{R^{2}}{(\operatorname*{Im}\xi)^{2}}\label{Poincare}%
\end{equation}

It seems natural to expect that LFT, at least in the semi-classical regime
$c_{L}>25$, also allows a solution corresponding to a quantization of this
geometry. In this paper we present what we believe might be the solution to
this problem. Surprisingly, we find an infinite set of different consistent
solutions parameterized by a couple of positive integers $(m,n)$ (which can be
put in natural correspondence with the degenerate representations of the
Virasoro algebra). Only the set $(1,n)$ has a smooth behavior as
$b\rightarrow0$ and therefore can be literally called the ``quantization'' of
(\ref{psphere}). Remarkably, all the solutions of this $(1,n)$ series are
indistinguishable in the classical limit (and even at the one-loop level),
dependence on $n$ appearing only in two-loop corrections. However, actual
higher loop calculations show that only the solution $(m,n)=(1,1)$ is
consistent with the standard loop perturbation theory. Therefore we are
inclined to interpret this last solution as the ``basic'' one, corresponding
to a ``natural'' quantization of the Lobachevskiy plane. Although the nature
of other solutions is still beyond our understanding (even of the
``perturbative'' series $(1,n)$, $n>1$) they probably can be speculated as
describing different phases of quantum gravity.

In principle all local properties of a field theory are encoded in its
operator product expansions. The latter are basically known in LFT (see
\cite{DO, AAl}). To have a complete description we also need certain
information of what is happening ``faraway'' from the observer, i.e, about the
boundary conditions at infinity. This information is encoded in the wave
function of the state which ``comes from infinity'', the so-called out-vacuum.
In order, the out-vacuum wave function can be described as the set of vacuum
expectation values (VEV's) of all local fields in the theory. In conformal
field theory, like LFT, it suffices to determine the VEV's (or one-point
functions) of all primary operators. The basic Liouville primaries are the
exponential fields
\begin{equation}
V_{\alpha}=\exp(2\alpha\phi)\label{Va}%
\end{equation}
of dimensions $\Delta_{\alpha}=\alpha(Q-\alpha)$. Thus the set of VEV's
$\left\langle V_{\alpha}\right\rangle $ is just the complementary information
we need to describe LFT in the pseudosphere geometry. In this paper we mainly
concentrate on this characteristic.

The paper is arranged as follows. In sect.2 the bootstrap technique is applied
to derive the one-point functions $\left\langle V_{\alpha}\right\rangle $. We
observe that all the out-vacuums $(m,n)$, if considered as conformal boundary
conditions at absolute, allow only finite set of boundary operators. In
particular, the basic out-vacuum $(1,1)$ does not contain any boundary fields
except the identity operator and its conformal descendents. Few simplest
bulk-boundary structure constants are also derived in this section. Certain
properties of the solutions are discussed in section 3. This includes
perturbative expansions of the one-point functions and some evidence about the
content of the boundary operators in the state $(m,n)$. In sect.4 one- and
two-loop contributions to the one-point functions are evaluated in the
framework of standard Feynmann diagram technique. At two loops these
calculations agree with the expansion of the ``basic'' vacuum state $(1,1)$.

In sect.5 the powerful modular bootstrap technique is applied to verify the
consistency of the proposed operator content at the out-vacuum states $(m,n)$.
Partition function of an annulus with ``boundaries'' corresponding to
different out-vacua $(m,n)$ is considered. It turns out that the modular
invariance of this partition function perfectly agrees with the suggested
operator content and can be further implemented for ``finite'' boundary
conditions discussed in ref.\cite{Fateev}.

With a finite set of boundary operators any two-point function in the bulk is
constructed as a finite sum of four-point conformal blocks. In sect.6 we
develop this construction explicitly for two simplest vacua $(1,1)$ and
$(1,2)$ and verify numerically that it satisfies the bulk-boundary bootstrap.
Some outlook and discussion is presented in sect.7.

\section{One-point bootstrap}

The basic assumption of the further development is that the out-vacuum state
generated by the absolute of the pseudosphere is conformally invariant, i.e.,
consists of a superposition of the Ishibashi states \cite{Ishibashi}. The
one-point functions of primary fields are nothing but the amplitudes of
different Ishibashi primaries in the out-vacuum wave function.

In this section we will use the Poincar\'{e} model of the Lobachevskiy plane
with complex coordinate $\xi$ in the upper half plane. Due to the conformal
invariance the coordinate dependence of any one-point function is prescribed
by the dimension of the operator
\begin{equation}
\left\langle V_{\alpha}(\xi)\right\rangle =\frac{U(\alpha)}{\left|  \xi
-\bar{\xi}\right|  ^{2\Delta_{\alpha}}}\label{one-p}%
\end{equation}
Thus we will call coordinate independent function $U(\alpha)$ the one-point
function and normalize it in the usual in field theory way $U(0)=1$.

Of course, local properties of LFT do not depend on the boundary conditions.
In particular the set of (bulk) degenerate fields
\begin{equation}
\Phi_{m,n}=\exp\left(  ((1-m)b^{-1}+(1-n)b)\phi\right) \label{Phimn}%
\end{equation}
still exists for any pair of positive $m$ and $n$. Therefore one can make use
of the trick applied by J.Teschner in the study of the operator algebra
\cite{Teschner}\ (see also \cite{Gervais} for very similar discussions).
Consider the following auxiliary two-point correlation function with the
insertion of an operator $\Phi_{1,2}=V_{-b/2}$
\begin{equation}
G_{-b/2,\alpha}(\xi,\xi^{\prime})=\left\langle V_{-b/2}(\xi)V_{\alpha}%
(\xi^{\prime})\right\rangle \label{aux}%
\end{equation}
Degenerate fields have very special structure of the operator product
expansions. In particular, the product $\Phi_{1,2}V_{\alpha}$ contains in the
right hand side only two primary fields $V_{\alpha-b/2}$ and $V_{\alpha+b/2}$.
Function (\ref{aux}) is therefore combined of two degenerate conformal blocks
\begin{equation}
G_{-b/2,\alpha}=\frac{\left|  \xi^{\prime}-\bar{\xi}^{\prime}\right|
^{2\Delta_{\alpha}-2\Delta_{12}}}{\left|  \xi-\bar{\xi}^{\prime}\right|
^{4\Delta_{a}}}\left[  C_{+}(\alpha)U(\alpha-b/2)\mathcal{F}_{+}(\eta
)+C_{-}(\alpha)U(\alpha+b/2)\mathcal{F}_{-}(\eta)\right] \label{Upm}%
\end{equation}
In our normalization the special structure constants $C_{\pm}(\alpha)$ read
explicitly \cite{Teschner, Fateev}
\begin{align}
C_{+}(\alpha) &  =1\label{Cpm}\\
C_{-}(\alpha) &  =-\pi\mu\frac{\Gamma(2\alpha b-b^{2}-1)\Gamma(1-2\alpha
b)\Gamma(1+b^{2})}{\Gamma(2+b^{2}-2\alpha b)\Gamma(2\alpha b)\Gamma(-b^{2}%
)}\nonumber
\end{align}
Degenerate conformal blocks $\mathcal{F}_{\pm}(\eta)$ are functions of the
projective invariant
\begin{equation}
\eta=\frac{(\xi-\xi^{\prime})(\bar{\xi}-\bar{\xi}^{\prime})}{(\xi-\bar{\xi
}^{\prime})(\bar{\xi}-\xi^{\prime})}\label{eta}%
\end{equation}
They are known explicitly and can be expressed in terms of hypergeometric
functions
\begin{align}
\mathcal{F}_{+}(\eta) &  =\eta^{\alpha b}(1-\eta)^{-b^{2}/2}{}_{1}%
F_{2}(2\alpha b-2b^{2}-1,-b^{2},2\alpha b-b^{2},\eta)\label{FF}\\
\mathcal{F}_{-}(\eta) &  =\eta^{1+b^{2}-\alpha b}(1-\eta)^{-b^{2}/2}{}%
_{1}F_{2}(-b^{2},1-2\alpha b,2+b^{2}-2\alpha b,\eta)\nonumber
\end{align}

The same expression (\ref{Upm}) can be rewritten also in terms of the
cross-channel degenerate blocks $\mathcal{G}_{\pm}(\eta)$
\begin{equation}
G_{-b/2,\alpha}=\frac{\left|  \xi^{\prime}-\bar\xi^{\prime}\right|
^{2\Delta_{\alpha}-2\Delta_{12}}}{\left|  \xi-\bar\xi^{\prime}\right|
^{4\Delta_{a}}}\left[  B^{(+)}(\alpha)\mathcal{G}_{+}(\eta)+B^{(-)}%
(\alpha)\mathcal{G}_{-}(\eta)\right] \label{Rpm}%
\end{equation}
where
\begin{align}
\mathcal{G}_{+}(\eta) &  =\eta^{\alpha b}(1-\eta)^{-b^{2}/2}{}_{1}F_{2}%
(-b^{2},2\alpha b-2b^{2}-1,-2b^{2},1-\eta)\label{GF}\\
\mathcal{G}_{-}(\eta) &  =\eta^{\alpha b}(1-\eta)^{1+3b^{2}/2}{}_{1}%
F_{2}(1+b^{2},2\alpha b,2+2b^{2},1-\eta)\nonumber
\end{align}
The boundary structure constants $B^{(\pm)}(\alpha)$ can be determined from
the relations
\begin{align}
\mathcal{F}_{+}(\eta) &  =\frac{\Gamma(2\alpha b-b^{2})\Gamma(1+2b^{2}%
)}{\Gamma(1+b^{2})\Gamma(2\alpha b)}\mathcal{G}_{+}(\eta)+\frac{\Gamma(2\alpha
b-b^{2})\Gamma(-1-2b^{2})}{\Gamma(2\alpha b-2b^{2}-1)\Gamma(-b^{2}%
)}\mathcal{G}_{-}(\eta)\label{cross}\\
\mathcal{F}_{-}(\eta) &  =\frac{\Gamma(2+b^{2}-2\alpha b)\Gamma(1+2b^{2}%
)}{\Gamma(1+b^{2})\Gamma(2+2b^{2}-2\alpha b)}\mathcal{G}_{+}(\eta
)+\frac{\Gamma(2+b^{2}-2\alpha b)\Gamma(-1-2b^{2})}{\Gamma(1-2\alpha
b)\Gamma(-b^{2})}\mathcal{G}_{-}(\eta)\nonumber
\end{align}
The block $\mathcal{G}_{-}(\eta)$ is recognized as corresponding to the identity
boundary operator of dimension $0$ while the boundary dimension $\Delta
_{13}=-1-2b^{2}$ corresponding to the block $\mathcal{G}_{+}(\eta)$ suggests
to identify it as the contribution of the degenerate boundary operator
$\psi_{1,3}$.

Projective invariant (\ref{eta}) can be interpreted in terms of the geodesic
distance $s(\xi,\xi^{\prime})$ on the pseudosphere. In the classical metric
(\ref{Poincare})
\begin{equation}
\eta=\tanh^{2}\frac{s}{2R}\label{distance}%
\end{equation}
It is important that on the pseudosphere as $\eta\rightarrow1$ the geodesic
distance becomes infinite. In a unitary field theory a two-point correlation
function is expected to decay in a product of the one-point ones as the
distance goes to infinity. The corresponding contribution is provided by the
identity operator. Therefore in a unitary theory with the usual large-distance
decay of correlations one would expect for $B^{(-)}(\alpha)$%
\begin{equation}
B^{(-)}(\alpha)=U(\alpha)U(-b/2)\label{Decay}%
\end{equation}
Together with (\ref{cross}) this gives the following non-linear functional
equation for $U(\alpha)$%
\begin{equation}
\ \frac{\Gamma(-b^{2})U(\alpha)U(-b/2)}{\Gamma(-1-2b^{2})\Gamma(2\alpha
b-b^{2})}=\frac{U(\alpha-b/2)}{\Gamma(2\alpha b-2b^{2}-1)}-\frac{\pi\mu
\Gamma(1+b^{2})U(\alpha+b/2)}{(2\alpha b-b^{2}-1)\Gamma(-b^{2})\Gamma(2\alpha
b)}\label{UUU}%
\end{equation}

Of course this equation admits many solutions. The set of the solutions can be
restricted largely by adding a similar ``dual'' functional equation where
$\alpha$ is shifted in $b^{-1}/2$ instead of $b/2$ in (\ref{UUU}). Dual
equation arises from the same calculation as (\ref{UUU}) but with the
degenerate field $\Phi_{21}$ taken instead of $\Phi_{12}$ in the auxiliary
two-point function (\ref{aux}). Due to the duality of LFT (see e.g.,
\cite{AAl}) this amounts the substitution $b\rightarrow b^{-1}$,
$\mu\rightarrow\tilde{\mu}$ in (\ref{UUU}). Here
\begin{equation}
\pi\tilde{\mu}\gamma(b^{-2})=\left(  \pi\mu\gamma(b^{2})\right)  ^{1/b^{2}%
}\label{mumutwiddle}%
\end{equation}
and as usual $\gamma(x)=\Gamma(x)/\Gamma(1-x)$.

It seems that (at least for real incommensurable values of $b$ and $1/b$) all
possible solutions fall into an infinite family parameterized by two positive
integers $(m,n)$
\begin{equation}
U_{m,n}(\alpha)=\frac{\sin(\pi b^{-1}Q)\sin(\pi mb^{-1}(2\alpha-Q))}{\sin(\pi
mb^{-1}Q)\sin(\pi b^{-1}(2\alpha-Q))}\frac{\sin(\pi bQ)\sin(\pi nb(2\alpha
-Q))}{\sin(\pi nbQ)\sin(\pi b(2\alpha-Q))}U_{1,1}(\alpha)\label{Umn}%
\end{equation}
where the ``basic'' $(1,1)$ one-point function reads
\begin{equation}
U(\alpha)=U_{1,1}(\alpha)=\frac{\left[  \pi\mu\gamma(b^{2})\right]
^{-\alpha/b}\Gamma(bQ)\Gamma(Q/b)Q}{\Gamma(b(Q-2\alpha))\Gamma(b^{-1}%
(Q-2\alpha))(Q-2\alpha)}\label{U11}%
\end{equation}

In the next section we will discuss some properties of these
solutions. Now let's take a look at the contribution of the block
$\mathcal{G}_{-}(\eta)$ to (\ref{aux}). This term is interpreted as the
contribution of the boundary operator $\psi_{1,3}$. Combining (\ref{cross})
and (\ref{Umn}) one finds
\begin{align}
\frac{B_{m,n}^{(+)}(\alpha)}{U_{m,n}(\alpha)} &  =(-)^{m-1}\left[  \pi
\mu\gamma(b^{2})\right]  ^{1/2}\frac{\Gamma(1+2b^{2})\Gamma(1-2b\alpha
)\Gamma(2\alpha b-2b^{2}-1)}{\pi\Gamma(b^{2})}\label{Bmn}\\
&  \ \ \times\frac{\sin(2\pi nb(\alpha-b))\sin(2\pi b\alpha)-\sin(2\pi
nb\alpha)\sin(2\pi b(\alpha-b))}{\sin(\pi nb(2\alpha-b))}\nonumber
\end{align}
In the standard CFT picture $B_{m,n}^{(+)}(\alpha)$ is composed from the
bulk-boundary structure constants $R_{m,n}^{(1,3)}(\alpha)$ for the operators
$V_{\alpha}$ and $V_{-b/2}$ merging to the boundary operator $\psi_{13}$ near
the $(m,n)$ boundary
\begin{equation}
B_{m,n}^{(+)}(\alpha)=R_{m,n}^{(1,3)}(\alpha)R_{m,n}^{(1,3)}(-b/2)D_{m,n}%
^{(1,3)}\label{Rmn}%
\end{equation}
Here $D_{m,n}^{(1,3)}$ stands for the boundary two-point function of two
operators $\psi_{1,3}$.

In principle the bootstrap technique allows to continue this process and
calculate all bulk-boundary structure constants $R_{m,n}^{(p,q)}(\alpha)$
corresponding to any degenerate boundary operator with odd $p$ and $q$. Here
we will not proceed systematically along this line. What can be already seen
from eq.(\ref{Bmn}) is that $R_{m,n}^{(1,3)}(\alpha)$ vanishes for the basic
out-vacuum state $(m,n)=(1,1)$. The following guess (which will be further
supported in the subsequent sections) seems rather natural. \emph{The basic
vacuum }$(1,1)$ \emph{contains no primary boundary operators except the
identity}. In this sense the basic vacuum is similar to the basic conformal
boundary condition discovered by J.Cardy \cite{Cardy} in the context of
rational conformal field theories.

The whole variety of vacua $(m,n)$ in this picture is naturally associated
with the boundary conditions corresponding to the degenerate fields
(\ref{Phimn}) themselves. Then, the content of boundary operators acting on
the vacuum $(m,n)$ (or, more generally, of the juxtaposition operators between
different vacua $(m,n)$ and $(m^{\prime},n^{\prime})$) is determined by the
fusion algebra, exactly as in the rational case. For instance, the vacuum
$(1,2)$ contains only identity boundary operator ($\psi_{1,1}=I$) and the
degenerate field $\psi_{1,3}$.

In principle, all these suggestions can be verified by systematic calculations
of the higher structure constants. We choose to postpone this difficult
problem for future studies. Instead in sect.5 we will see that the above
pattern is perfectly consistent with the modular bootstrap of the annulus
partition function.

\section{The one-point function}

The solution (\ref{Umn}) for the one-point function bears some remarkable properties.

\textbf{1. One-point Liouville equation.} For all $(m,n)$%
\begin{equation}
U_{m,n}(b)=\frac Q{\pi\mu b}\label{Ub}%
\end{equation}
(of course the dual relation $U_{m,n}(1/b)=bQ/(\pi\tilde\mu)$ with $\tilde\mu$
from (\ref{mumutwiddle}) is also valid). In particular, this means that the
quantum Liouville equation in the form (\ref{Liouv}) holds on the one-point
level. Indeed, if the quantum Liouville field $\phi$ is defined as
\begin{equation}
\phi=\frac12\left.  \frac{\partial V_{\alpha}}{\partial\alpha}\right|
_{\alpha=0}\label{phi}%
\end{equation}
it follows from (\ref{one-p}) that (we take the Poincar\'e model
(\ref{Poincare}) for the moment)
\begin{equation}
\left\langle \phi(\xi)\right\rangle _{m,n}=-Q\log\left|  \xi-\bar\xi\right|
^{2}+\left.  \partial U_{m,n}(\alpha)/\partial\alpha\right|  _{\alpha
=0}\label{<phi>}%
\end{equation}
and therefore
\begin{equation}
\Delta\left\langle \phi(\xi)\right\rangle _{m,n}=\frac{4Q}{\left|  \xi-\bar
\xi\right|  ^{2}}=4\pi\mu b\left\langle V_{b}(\xi)\right\rangle _{m,n}%
\label{Vb}%
\end{equation}

\textbf{2. Normalization. }All the one-point functions $U_{m,n}(\alpha)$ are
normalized by $U_{m,n}(0)=1$ so that the expectation value of the identity
operator is $1$. It will prove convenient to introduce the function
\begin{equation}
W(\lambda)=\frac{2\pi\lambda\left(  \pi\mu\gamma(b^{2})\right)  ^{-\lambda/b}%
}{\Gamma(1-2\lambda/b)\Gamma(1-2b\lambda)}\label{W}%
\end{equation}
It satisfies the following functional relations
\begin{align}
W(\lambda)W(-\lambda) &  =-\sin(2\pi b\lambda)\sin(2\pi\lambda/b)\label{WW}\\
\frac{W(iP)}{W(-iP)} &  =S_{L}(P)\nonumber
\end{align}
$S_{L}(P)$ being the standard Liouville reflection amplitude \cite{AAl}
\begin{equation}
S_{L}(P)=\left(  \pi\mu\gamma(b^{2})\right)  ^{-2iP/b}\frac{\Gamma
(1+2iP/b)\Gamma(1+2ibP)}{\Gamma(1-2iP/b)\Gamma(1-2ibP)}\label{SL}%
\end{equation}
Notice, that $U(Q/2+\lambda)$ differs from $W(\lambda)$ only in overall
normalization
\begin{equation}
U(\alpha)=\frac{W(\alpha-Q/2)}{W(-Q/2)}\label{UW}%
\end{equation}

\textbf{3. Reflection relation. }Apparently, all the solutions $U_{m,n}%
(\alpha)$ satisfy the so-called reflection relations (see e.g. ref. \cite{AAl}
for details)
\begin{equation}
U_{m,n}(\alpha)=S_{L}\left(  \frac{2\alpha-Q}{2i}\right)  U_{m,n}%
(Q-\alpha)\label{R}%
\end{equation}
with the Liouville reflection amplitude (\ref{SL}). This is consistent with
the local properties of the LFT\ primary fields suggested in ref.\cite{AAl}.

\textbf{4. In the ``basic'' vacuum }$\mathbf{(m,n)=(1,1)}$\textbf{\ }the
one-point function $U_{1,1}(\alpha)$ reads
\begin{equation}
U_{1,1}(\alpha)=U(\alpha)=\frac{\left[  \pi\mu\gamma(b^{2})\right]
^{-\alpha/b}\Gamma(2+b^{2})\Gamma(1+1/b^{2})}{\Gamma(2+b^{2}-2b\alpha
)\Gamma(1+b^{-2}-2\alpha/b)}\label{u11}%
\end{equation}
while from eq.(\ref{Bmn}) we have
\begin{equation}
B_{1,1}^{(+)}(\alpha)=0\label{zero}%
\end{equation}
i.e., as it has been mentioned in the previous section, boundary field
$\psi_{1,3}$ does not contribute to (\ref{aux}) in the basic vacuum. As it has
been suggested above, this is a particular instance of a more general
phenomenon: In the basic vacuum the only primary boundary operator is the
identity one$.$ This feature makes the basic state rather distinguished among
the whole set (\ref{Umn}). We are inclined to identify it as the generic
out-vacuum state of LFT on the Lobachevskiy plane. In the next section this
statement is checked against the ordinary (loop) perturbation theory up to
two-loop order.

\textbf{5. Perturbative expansions. }If $m>1$ the one-point function
$U_{m,n}(\alpha)$ is essentially singular at $b\rightarrow0$ and therefore
admits no usual classical limit. Singular behavior in the classical limit
makes it rather difficult to interpret these states as a ``quantization'' of
any classical metric. For the moment let us restrict attention to the case
$m=1$ where $U_{1,n}(\alpha)$ is smooth in the classical limit and can be
expanded in an (asymptotic) series in $b$.

Take the disk model (\ref{psphere}) of the pseudosphere, where the one-point
function reads
\begin{equation}
\left\langle V_{\alpha}(z)\right\rangle =\frac{U(\alpha)}{(1-z\bar
{z})^{2\alpha(Q-\alpha)}}\label{Uz}%
\end{equation}
First, it is convenient to take the logarithm of this function and expand it
in powers of $\alpha$
\begin{equation}
\log\left\langle V_{\alpha}(z)\right\rangle =\sum_{k=1}^{\infty}\frac
{(2\alpha)^{k}}{k!}G_{k}(b^{2})\label{UG}%
\end{equation}
The coefficients are readily interpreted as the VEV's of ``connected powers''
of the Liouville field $\phi$, e.g.
\begin{align}
G_{1}(b) &  =\left\langle \phi\right\rangle \nonumber\\
G_{2}(b) &  =\left\langle \phi^{2}\right\rangle -\left\langle \phi
\right\rangle ^{2}\label{Gn}\\
G_{3}(b) &  =\left\langle \phi^{3}\right\rangle -3\left\langle \phi
\right\rangle \left\langle \phi\right\rangle ^{2}+2\left\langle \phi
\right\rangle ^{3}\nonumber\\
&  \ \text{etc.}\nonumber
\end{align}
In order, each $G_{k}(b^{2})$ allows an asymptotic expansion in powers of $b$%
\begin{equation}
G_{k}(b^{2})\sim\sum_{l=k-1}^{\infty}G_{k}^{(l)}b^{2l-k}\label{Loop}%
\end{equation}
For the first few coefficients we find (for the general ``perturbative''
solution $U_{1,n}(\alpha)$)
\begin{align}
G_{1}(b) &  \sim-\frac{1}{2b}\log\left(  \pi\mu b^{2}(1-z\bar{z})^{2}\right)
+b\left(  -\log(1-z\bar{z})+\frac{3}{2}\right) \nonumber\\
&  +b^{3}\left(  \frac{\pi^{2}}{6}-\frac{13}{12}+\frac{\pi^{2}(n^{2}-1)}%
{3}\right)  +\ldots\label{GG}\\
G_{2}(b) &  \sim\log(1-z\bar{z})-1+b^{2}\left(  \frac{3}{2}-\frac{\pi^{2}}%
{6}-\frac{\pi^{2}(n^{2}-1)}{3}\right)  +\ldots\nonumber\\
G_{3}(b) &  \sim-b+\ldots\nonumber
\end{align}
Notice that the classical limit $G_{1}^{(0)}$ and one-loop terms 
$G_{1}^{(1)}$ and $G_{2}^{(1)}$ are completely unsensitive to the sort of 
the out-vacuum $(1,n)$, the dependence on the vacuum 
appearing only in the terms corresponding to two and higher loops ($l\geq2$).

\section{Loop perturbation theory}

Expansions (\ref{GG}) can be compared against the standard loop perturbation
theory around the classical solution (\ref{psphere}). In this section we will
use the most ``naive'' perturbation theory which treats the LFT action
\begin{equation}
A_{L}=\int\left[  \frac1{4\pi}(\partial_{a}\phi)^{2}+\mu e^{2b\phi}\right]
d^{2}x\label{AL}%
\end{equation}
straightforwardly and leads to a diagram technique with different tadpole
diagrams. In this technique the standard Liouville scaling exponents do not
appear exactly from the very beginning but are the result of complete
summation (however rather simple) of the tadpoles. Of course, more
sophisticated diagram techniques can be developed for the Liouville field
theory, which automatically take advantage of the Weil and conformal
invariance of the theory to reduce the set of tadpole diagrams and start with
the exact exponents \cite{Thorn}. These versions of LFT\ perturbation theory
(which are of course equivalent to the ``naive'' one) have essential
advantages at higher loop calculations where tadpole diagrams are rather
numerous and their counting becomes a certain combinatorial problem.

The Weil invariance of LFT means that the theory with the action (\ref{AL}) is
equivalent to LFT in any background metric $g_{ab}(x)$. In general the LFT
action reads
\begin{equation}
A_{L}[g]=\int\left[  \frac{1}{4\pi}g^{ab}\partial_{a}\phi\partial_{b}%
\phi+\frac{Q}{4\pi}R\phi+\mu e^{2b\phi}\right]  \sqrt{g}d^{2}x\label{ALg}%
\end{equation}
where $R$ is the scalar curvature of the background metric, the theory being
essentially independent on $g$. ``Naive'' form (\ref{AL}) of the Liouville
action implies the ``trivial'' background metric $g_{ab}=\delta_{ab}$ in the
parametric space, e.g. inside the unit disk (\ref{psphere}). This means that
all ultraviolet divergencies are regularized with respect to this trivial metric.

Of course the classical solution $\phi_{\text{cl}}$%
\begin{equation}
e^{2b\phi_{\text{cl}}}=\frac1{\pi\mu b^{2}(1-z\bar z)^{2}}\label{phicl}%
\end{equation}
does not depend on any background metric. Substituting
\begin{equation}
\phi=\phi_{\text{cl}}+\chi\label{chi}%
\end{equation}
into (\ref{AL}) we have
\begin{equation}
A_{L}=A_{L}^{(\text{cl})}+\int\left[  \frac1{4\pi}(\partial_{a}\chi)^{2}%
+\frac{e^{2b\chi}-2b\chi-1}{\pi b^{2}(1-z\bar z)^{2}}\right]  d^{2}%
x\label{Achi}%
\end{equation}

At the classical (zero-loop) level only $G_{1}$ is non-zero
\begin{equation}
G_{1}=-\frac1{2b}\log\left(  \pi\mu b^{2}(1-z\bar z)^{2}\right)
+\ldots\label{Gcl}%
\end{equation}
and consistent with $G_{1}^{(0)}$ in expansion (\ref{GG}). The one-loop
(Gaussian) part of the action (\ref{Achi}) has the form
\begin{equation}
\frac1{2\pi}\int\left[  \frac12(\partial_{a}\chi)^{2}+\frac{4\chi^{2}%
}{(1-z\bar z)^{2}}\right]  d^{2}x\label{Gauss}%
\end{equation}
It leads to the following ``bare'' propagator of the field $\chi$
\begin{equation}
g(z,z^{\prime})=\left\langle \chi(z,\bar z)\chi(z^{\prime},\bar z^{\prime
})\right\rangle =-\frac12\left(  \frac{1+\eta}{1-\eta}\log\eta+2\right)
\label{prop}%
\end{equation}
The propagator depends only on the invariant
\begin{equation}
\eta=\frac{(z-z^{\prime})(\bar z-\bar z^{\prime})}{(1-z\bar z^{\prime})(1-\bar
zz^{\prime})}\label{etaz}%
\end{equation}
which is related to the ``geodesic distance'' $s$ between the points $z$ and
$z^{\prime}$ as in eq.(\ref{distance})

\begin{figure}
[tbh]
\begin{center}
\includegraphics[
height=3.4947in,
width=2.9836in
]%
{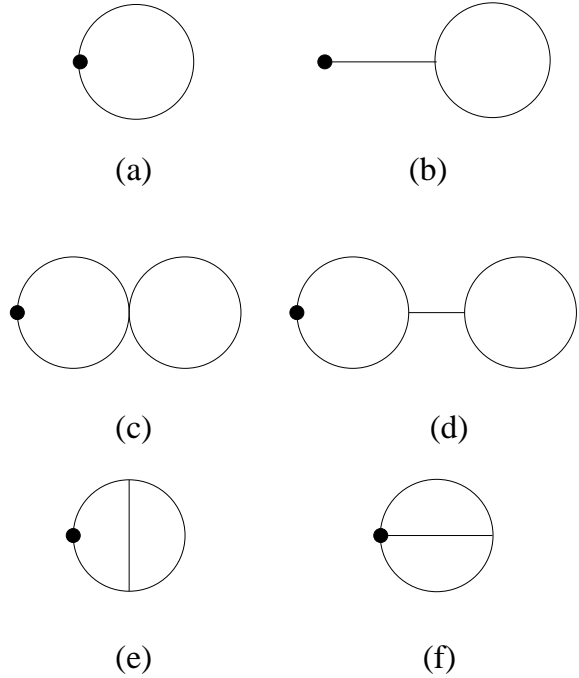}%
\caption{Diagrams contributing to the expectation values $G_{1}$, $G_{2}$ and
$G_{3}$ at one- and two-loop order.}%
\label{diag}%
\end{center}
\end{figure}

The simplest one-loop diagram of fig.\ref{diag}a contributes to $\left\langle
\chi^{2}(z,\bar z)\right\rangle $%
\begin{equation}
\text{\textbf{Fig.\ref{diag}a}}=\lim_{z^{\prime}\rightarrow z}\left(
g(z,z^{\prime})+\log\left|  z-z^{\prime}\right|  \right)  =\log(1-z\bar
z)-1\label{tadpole2}%
\end{equation}
With this result it is easy to evaluate the one-loop correction to
$\left\langle \chi(z,\bar z)\right\rangle $ as given by the diagram
Fig.\ref{diag}b
\begin{align}
\text{\textbf{Fig.\ref{diag}b}} &  =-4b\int g(z,z^{\prime})\frac{\left\langle
\chi^{2}(z^{\prime})\right\rangle }{(1-z^{\prime}\bar z^{\prime})^{2}}%
d^{2}z^{\prime}\label{tadpole1}\\
&  =b\left(  -\log(1-z\bar z)+3/2)\right) \nonumber
\end{align}
Both (\ref{tadpole2}) and (\ref{tadpole1}) agree with the one-loop terms in
(\ref{GG}).

In general there is no need to calculate separately the contribution to
$G_{1}$ at any loop order once the corresponding contributions to higher $G$'s
are known. This is due to the following Ward identity
\begin{equation}
\left\langle \chi\right\rangle =\frac2{\pi b}\int g(z,z^{\prime}%
)\frac{\left\langle e^{2b\chi(z^{\prime})}\right\rangle -2b\left\langle
\chi(z^{\prime})\right\rangle -1}{(1-z^{\prime}\bar z^{\prime})^{2}}%
d^{2}z^{\prime}\label{Ward}%
\end{equation}
which apparently holds order by order in the loop perturbation theory. Notice
that this identity can be considered as a perturbative equivalent of the exact
relation (\ref{Ub}).%

The simplest two-loop correction is the leading contribution to $G_{3}.$ There
is only one diagram Fig.\ref{diag}f which is readily evaluated
\begin{equation}
\text{\textbf{Fig.\ref{diag}f}}=-\frac{8b}\pi\int\frac{g^{3}(z,z^{\prime}%
)}{(1-z^{\prime}\bar z^{\prime})^{2}}d^{2}z^{\prime}=-8b\int_{0}^{1}%
\frac{g^{3}(\eta)d\eta}{(1-\eta)^{2}}=-b\label{Figf}%
\end{equation}
again in agreement with the exact value of $G_{3}^{(2)}$ in (\ref{GG}). Next,
take the two-loop corrections to $G_{2}$. Let us first evaluate the tadpole
contributions of Fig.\ref{diag}c and Fig.\ref{diag}d. Since
\begin{equation}
b\left(  \text{\textbf{Fig.\ref{diag}a}}\right)  +\text{\textbf{Fig.\ref{diag}%
b}}=b/2\label{sum}%
\end{equation}
we have
\begin{align}
\text{\textbf{Fig.\ref{diag}c}}+\text{\textbf{Fig.\ref{diag}d}} &
=-4b^{2}\int_{0}^{1}\frac{g^{2}(\eta)}{(1-\eta)^{2}}d\eta\label{tadpole3}\\
&  =\frac{6-\pi^{2}}9b^{2}\nonumber
\end{align}
To evaluate the two-loop diagram Fig.\ref{diag}e we need the following
two-point function
\begin{equation}
g_{1,2}(\eta)=\left\langle \chi(z)\chi^{2}(z^{\prime})\right\rangle
\label{g12}%
\end{equation}
as given by the following one-loop diagram
\begin{align}
g_{1,2}(z,z^{\prime}) &  =%
\raisebox{-0.2413in}{\includegraphics[
height=0.6019in,
width=1.5333in
]%
{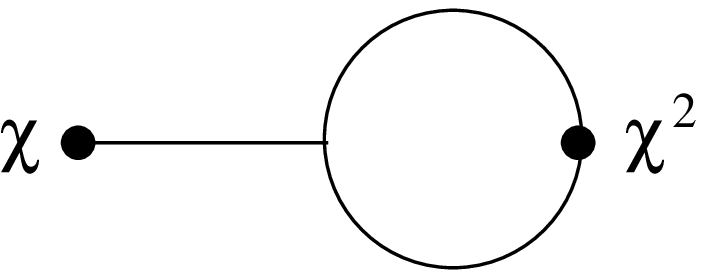}%
}%
=\int\frac{g(z,z^{\prime\prime})g^{2}(z^{\prime\prime},z^{\prime}%
)}{(1-z^{\prime\prime}\bar z^{\prime\prime})^{2}}d^{2}z^{\prime\prime
}\label{g121}\\
&  =-\frac\pi8\left(  \frac{\eta\log^{2}\eta}{(1-\eta)^{2}}-1\right) \nonumber
\end{align}
For the two-loop diagram we obtain
\begin{align}
\text{\textbf{Fig.\ref{diag}e}} &  =\frac{32b^{2}}{\pi^{2}}\int\frac
{g_{1,2}(z,z^{\prime})g(z,z^{\prime})}{(1-z^{\prime}\bar z^{\prime})^{2}}%
d^{2}z^{\prime}\label{fige}\\
&  =\frac{15-\pi^{2}}{18}b^{2}\nonumber
\end{align}
Adding all the two-loop diagrams together results in
\begin{equation}
G_{2}^{(2)}=\text{\textbf{Fig.\ref{diag}c}}+\text{\textbf{Fig.\ref{diag}d}%
}+\text{\textbf{Fig.\ref{diag}e}}=\frac{9-\pi^{2}}6b^{2}\label{G22}%
\end{equation}
and agrees with the expansion (\ref{GG}) if we take $n=1$, i.e., for the
``basic'' vacuum state.

Here we will not develop further the loop perturbation theory for LFT on the
Lobachevskiy plane. To go at higher loop diagrammatic calculations it is worth
first to improve the technique to better handle the tadpole diagrams (which
become rather numerous at higher orders) and second to take advantage of the
space-time symmetries (the $SL(2,R)$ group) of the theory. We hope to turn at
these interesting points in close future.

\section{Modular bootstrap}

General non-degenerate Virasoro character is written as
\begin{equation}
\chi_{P}(\tau)=\frac{q^{P^{2}}}{\eta(\tau)}\label{chiP}%
\end{equation}
where
\begin{equation}
\eta\left(  \tau\right)  =q^{1/24}\prod_{n=1}^{\infty}(1-q^{n}%
),\;\;\;\;\;q=\exp(2i\pi\tau)\label{dedekind}%
\end{equation}
Here $P$ is related to the central charge and the dimension of the
representation via eqs.(\ref{cL}) and
\begin{equation}
\Delta_{P}=Q^{2}/4+P^{2}\label{DP}%
\end{equation}
Degenerate representations appear at \cite{Kac}
\begin{equation}
\Delta_{m,n}=Q^{2}/4-(m/b+nb)^{2}/4\label{Dmn}%
\end{equation}
where $(m,n)$ are positive integers. At general $b$ there is only one
null-vector at the level $mn$. Hence the degenerate character reads simply as
\begin{equation}
\chi_{m,n}(\tau)=\frac{q^{-(m/b+nb)^{2}/4}-q^{-(m/b-nb)^{2}/4}}{\eta(\tau
)}\label{chimn}%
\end{equation}
Applying the identity
\begin{align}
\chi_{P}(\tau^{\prime}) &  =e^{2i\pi P^{2}\tau^{\prime}}\sqrt{-i\tau^{\prime}%
}\eta^{-1}(\tau)\nonumber\\
\  &  =\sqrt{2}\int\chi_{P^{\prime}}(\tau)e^{4i\pi PP^{\prime}}dP^{\prime
}\label{root2}%
\end{align}
where
\begin{align}
\tau^{\prime} &  =-1/\tau\label{S}\\
q^{\prime} &  =\exp(2i\pi\tau^{\prime})\nonumber
\end{align}
we find
\begin{align}
\chi_{m,n}(\tau^{\prime}) &  =\sqrt{2}\int\chi_{P}(\tau)\left(  \cosh
2\pi(m/b+nb)P-\cos2\pi(m/b-nb)P\right)  dP\nonumber\\
\  &  =2\sqrt{2}\int\chi_{P}(\tau)\sinh(2\pi mP/b)\sinh(2\pi
nbP)dP\label{shmshn}%
\end{align}
In particular
\begin{align}
\chi_{1,1}(q^{\prime}) &  =2\sqrt{2}\int\chi_{P}(q)\sinh(2\pi bP)\sinh(2\pi
P/b)dP\label{sh11}\\
\  &  =\int\Psi_{1,1}(P)\Psi_{1,1}(-P)\chi_{P}(q)dP\nonumber
\end{align}
where we have set
\begin{align}
\Psi_{1,1}(P) &  =\frac{2^{3/4}2i\pi P}{\Gamma(1-2ibP)\Gamma(1-2iP/b)}(\pi
\mu\gamma(b^{2}))^{-iP/b}\label{Psi11}\\
\  &  =2^{3/4}W(iP)\nonumber
\end{align}
The function $W(\lambda)$ has been defined in eq.(\ref{W}). $\Psi_{1,1}(P)$ is
interpreted as the wave function of the basic out-vacuum state
\begin{equation}
\left\langle (1,1)\;\text{outvac}\right|  =\int\Psi_{1,1}(P)\left\langle
P\right|  dP\label{11outvac}%
\end{equation}
Here $\left\langle P\right|  $ are the Ishibashi states \cite{Ishibashi}
\begin{equation}
\left\langle P\right|  =\left\langle v_{P}\right|  \left(  1+\frac{L_{1}\bar
L_{1}}{2\Delta_{P}}+\ldots\right) \label{ishibashi}%
\end{equation}
for different primary states $v_{P}$. The last are assumed to be normalized
as follows
\begin{equation}
\left\langle v_{P}|v_{P^{\prime}}\right\rangle =\delta(P-P^{\prime
})\label{norm}%
\end{equation}

Let us now take (\ref{shmshn}) and represent it in the form
\begin{equation}
\chi_{m,n}(q^{\prime})=\int\Psi_{m,n}(P)\Psi_{1,1}(-P)\chi_{P}%
(q)dP\label{psimnpsi11}%
\end{equation}
with
\begin{equation}
\Psi_{m,n}(P)=\Psi_{1,1}(P)\frac{\sinh(2\pi mP/b)\sinh(2\pi nbP)}{\sinh(2\pi
P/b)\sinh(2\pi bP)}\label{Psimn}%
\end{equation}
(compare this expression with eq.(\ref{Umn})). This is naturally interpreted
as the wave function of a general $(m,n)$ out-vacuum state. It remains us to
verify the operator content in the decomposition of the ``partition
function''
\begin{align}
Z_{(m,n),(m^{\prime},n^{\prime})}(q) &  =\int\Psi_{m,n}(P)\Psi_{m^{\prime
},n^{\prime}}(-P)\chi_{P}(q)dP\label{Zmnmn}\\
\  &  =\int\frac{\sinh(2\pi mP/b)\sinh(2\pi nbP)\sinh(2\pi m^{\prime}%
P/b)\sinh(2\pi n^{\prime}bP)}{\sinh(2\pi P/b)\sinh(2\pi bP)}\chi
_{P}(q)dP\nonumber
\end{align}
Thanks to the identity
\begin{equation}
\sinh(2\pi nbP)\sinh(2\pi n^{\prime}bP)=\sum_{l=0}^{\min(n,n^{\prime})-1}%
\sinh(2\pi bP)\sinh(2\pi b(n+n^{\prime}-2l-1)P)\label{trig}%
\end{equation}
this results in the standard character set
\begin{equation}
Z_{(m,n),(m^{\prime},n^{\prime})}(q)=\sum_{k=0}^{\min(m,m^{\prime})-1}%
\sum_{l=0}^{\min(n,n^{\prime})-1}\chi_{m+m^{\prime}-2k-1,n+n^{\prime}%
-2l-1}(q')\label{standard}%
\end{equation}
determined by the fusion algebra of the degenerate representations.

Consider also a general non-degenerate character with $P=s/2$ (i.e., with
$\Delta=Q^{2}/4+s^{2}/4$)
\begin{equation}
\chi_{s/2}(q^{\prime})=\sqrt{2}\int\chi_{P}(q)\cos(2\pi sP)dP\label{chis}%
\end{equation}
This can be interpreted as
\begin{equation}
\chi_{s/2}(q^{\prime})=\int\Psi_{1,1}(P)\Psi_{s}(-P)\chi_{P}%
(q)dP\label{psispsi11}%
\end{equation}
if
\begin{align}
\Psi_{s}(P) &  =\frac{2^{-1/4}\Gamma(1+2ibP)\Gamma(1+2iP/b)\cos(2\pi
sP)}{-2i\pi P}(\pi\mu\gamma(b^{2}))^{-iP/b}\label{Psis}\\
&  =\frac{2^{-1/4}W(iP)\cos(2\pi sP)}{\sinh(2\pi bP)\sinh(2\pi P/b)}\nonumber
\end{align}
This wave function has been already discussed in ref.\cite{Fateev} (see also
\cite{Tshit} for modular considerations) in connection with certain local 
conformally
invariant boundary conditions in boundary LFT. Therefore it is natural to
associate such boundary state with a general non-degenerate
representation with $P=s/2$.

Next, let us decompose the following overlap integral
\begin{equation}
\int\Psi_{m,n}(P)\Psi_{s}(-P)\chi_{P}(q)dP=\sqrt{2}\int\chi_{P}(q)\frac
{\sinh(2\pi mP/b)\sinh(2\pi nbP)}{\sinh(2\pi P/b)\sinh(2\pi bP)}\cos(2\pi
sP)dP\label{overlap}%
\end{equation}
Again, we use the identity
\begin{equation}
\frac{\sinh(2\pi nbP)}{\sinh(2\pi bP)}=\sum_{l=1-n,2}^{n-1}\exp(2\pi
lbP)\label{trig2}%
\end{equation}
(here $\sum_{l=1-n,2}^{n-1}$ denotes the sum over the set
$l=\{-n+1,-n+3,\ldots,n-1\}$) to obtain
\begin{equation}
\int\Psi_{m,n}(P)\Psi_{s}(-P)\chi_{P}(q)dP=\sum_{k=1-m,2}^{m-1}\sum
_{l=1-n,2}^{n-1}\chi_{(s+i(k/b+lb))/2}(q^{\prime})\label{PsimnPsis}%
\end{equation}
i.e., the standard fusion of the degenerate representation $(m,n)$ and a
general one with $P=s/2$.

It remains us to analyze the partition function with two boundary conditions
characterized by different boundary parameters $s$ and $s^{\prime}$. It is
given by the overlap integral
\begin{align}
Z_{s,s^{\prime}} &  =\int\Psi_{s}(-P)\Psi_{s^{\prime}}(P)\chi_{P}%
(q)dP\nonumber\\
&  =\sqrt{2}\int\Psi_{s}(-P)\Psi_{s^{\prime}}(P)e^{-4i\pi PP^{\prime}}%
\chi_{P^{\prime}}(q^{\prime})dPdP^{\prime}\label{ss}\\
&  =\int_{0}^{\infty}\rho(P^{\prime})\chi_{P^{\prime}}(q^{\prime})dP^{\prime}\nonumber
\end{align}
where, according to \cite{Tshit} the density of states $\rho(P^{\prime})$ flowing along
the strip reads
\begin{align}
\rho(P^{\prime}) &  =2\sqrt{2}\int\Psi_{s}(-P)\Psi_{s^{\prime}}(P)e^{-4i\pi
PP^{\prime}}dP\label{rho}\\
&  =\int_{-\infty}^{\infty}\frac{2\cos(st)\cos(s^{\prime}t)}{\sinh
(bt)\sinh(t/b)}e^{-2iP^{\prime}t}\frac{dt}{2\pi}\nonumber
\end{align}
In this integral some regularization of the singularity at $P=0$ is implied.
Eq.(\ref{rho}) has to be compared with the logarithmic derivative
\begin{equation}
\rho(P)=-\frac i{2\pi}\frac d{dP}\log D_{B}(P|s,s^{\prime})\label{DB}%
\end{equation}
of the boundary Liouville two-point function constructed in \cite{Fateev}. As
it has been mentioned in \cite{Tshit} the two expressions match up to an
$s$-independent quantity. There is also some specific $s$-independent (but
still $P$-dependent) part of the two-point function, and consequently of the
density of states on the strip, which cannot be restored from the integral
(\ref{rho}) before the regularization at $P=0$ is specified.%

\begin{figure}
[tbh]
\begin{center}
\includegraphics[
height=3.3494in,
width=3.4584in
]%
{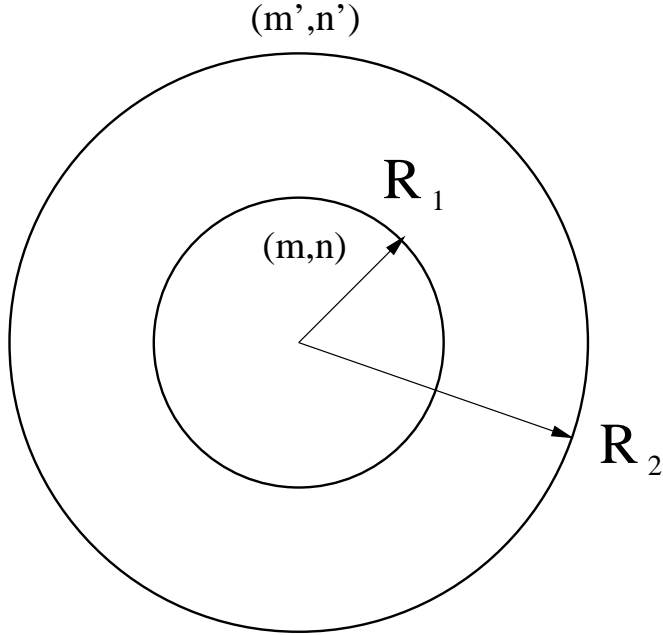}%
\caption{The annulus with two ``boundary conditions'' corresponding to the
out-vacuum states $(m,n)$ and $(m^{\prime},n^{\prime})$.}%
\label{annulus}%
\end{center}
\end{figure}

All the above calculations are quite formal. The underlying physical picture
involves LFT on an annulus with a (purely imaginary) modular parameter
\begin{equation}
\tau=\frac{i}{\pi}\log\frac{R_{2}}{R_{1}}\label{tau}%
\end{equation}
(see fig.\ref{annulus}). An annulus with the out-vacuum states at both
boundaries is interpreted as finite-temperature partition function of
gravitational modes in $AdS_2$ geometry \cite{Str}. Thus, the right-hand
side of (\ref{standard}) exposes the state content of such theory. Note
that in the case when both of the out-vacua are of $(1,1)$ type, the 
corresponding space of states contains only identity operator (i.e. the
$SL(2,R)$ invariant state found in \cite{Jackiw}) and its conformal
descendents. The situation is more difficult to interpret when the 
out-vacuum associated with one (or both) of the boundaries is of
$(m,n) \neq (1,1)$ type, in which cases (\ref{psimnpsi11}) and 
(\ref{standard}) indicate presence of nontrivial primary states with Kac 
dimensions (\ref{Dmn}). Proper interpretation of these states (and the
``excited'' out-vacua $(m,n) \neq (1,1)$ themselves) is one of the most 
interesting questions remaining open. Nevertheless, one can notice that,
at least on the formal level, the above modular pattern is strikingly
similar to the situation in boundary rational conformal field theories,
as discussed in \cite{Cardy}. Much of the similarity remains there when
one of the out-vacua is replaced by a local boundary condition $\Psi_s$;
the right-hand side of (\ref{PsimnPsis}) reveals the state content of 
``semi-infinite'' $AdS_2$, with one ``$AdS$ boundaries'' replaced by a local
boundary condition of \cite{Fateev} at a finite distance. Again, true
interpretation of these states still needs to be clarified.

\section{Boundary bootstrap}

With finite number of boundary fields any two-point function of bulk primary
fields is combined of finite number of conformal blocks. In this section we
use the structure constants calculated in sect.2 to construct this two-point
function in some simplest cases and verify that it satisfies the boundary
bootstrap relations.

We consider the general two-point function
\begin{equation}
G_{\alpha_{1}\alpha_{2}}(\xi_{1},\xi_{2})=\left\langle V_{\alpha_{1}}(\xi
_{1})V_{\alpha_{2}}(\xi_{2})\right\rangle _{m,n}\label{2p}%
\end{equation}
computed in some out-vacuum $(m,n)$. Let us study two simplest cases.

\textbf{1. ``Basic'' vacuum }$(\mathbf{m,n)=(1,1)}$\textbf{.} In the basic
out-vacuum $(m,n)=(1,1)$ there is only identity operator at the boundary. The
function reads
\begin{equation}
G_{\alpha_{1}\alpha_{2}}(\xi_{1},\xi_{2})=\frac{\left|  \xi_{2}-\bar{\xi}%
_{2}\right|  ^{2\Delta_{1}-2\Delta_{2}}U_{1,1}(\alpha_{1})U_{1,1}(\alpha_{2}%
)}{\left|  \xi_{1}-\bar{\xi}_{2}\right|  ^{4\Delta_{2}}}\mathcal{F}\left(
\begin{array}
[c]{cc}%
\alpha_{1} & \alpha_{2}\\
\alpha_{1} & \alpha_{2}%
\end{array}
,iQ/2,1-\eta\right) \label{11block}%
\end{equation}
where $\mathcal{F}$ is the standard four-point conformal block with
``intermediate'' dimension $\Delta=0$. To get rid of excessive multipliers it
is convenient to define a ``normalized'' two-point function as
\begin{equation}
g_{\alpha_{1}\alpha_{2}}(\eta)=\frac{\left\langle V_{\alpha_{1}}(\xi
_{1})V_{\alpha_{2}}(\xi_{2})\right\rangle _{1,1}}{\left\langle V_{\alpha_{1}%
}(\xi_{1})\right\rangle _{1,1}\left\langle V_{\alpha_{2}}(\xi_{2}%
)\right\rangle _{1,1}}\label{gaa}%
\end{equation}
which is simply expressed in terms of this single block
\begin{equation}
g_{\alpha_{1}\alpha_{2}}(\eta)=(1-\eta)^{2\Delta_{1}}\mathcal{F}\left(
\begin{array}
[c]{cc}%
\alpha_{1} & \alpha_{2}\\
\alpha_{1} & \alpha_{2}%
\end{array}
,iQ/2,1-\eta\right) \label{F}%
\end{equation}
and depends only on the invariant $\eta$.

Another representation of this function comes from the bulk operator product
expansion of the fields $V_{\alpha_{1}}(\xi_{1})$ and $V_{\alpha_{2}}(\xi
_{2})$. This gives the ``normalized'' two-point function (up to possible
discrete terms) in the form
\begin{equation}
g_{\alpha_{1}\alpha_{2}}(\eta)=(1-\eta)^{2\Delta_{1}}%
{\displaystyle\int_{-\infty}^{\infty}}
\frac{dP}{4\pi}\frac{C(\alpha_{1},\alpha_{2},Q/2+iP)U_{1,1}(Q/2-iP)}%
{U_{1,1}(\alpha_{1})U_{1,1}(\alpha_{2})}\mathcal{F}\left(
\begin{array}
[c]{cc}%
\alpha_{1} & \alpha_{1}\\
\alpha_{2} & \alpha_{2}%
\end{array}
,P,\eta\right) \label{Ft}%
\end{equation}%

\begin{figure}
[tbh]
\begin{center}
\includegraphics[
height=4.2497in,
width=4.9493in
]%
{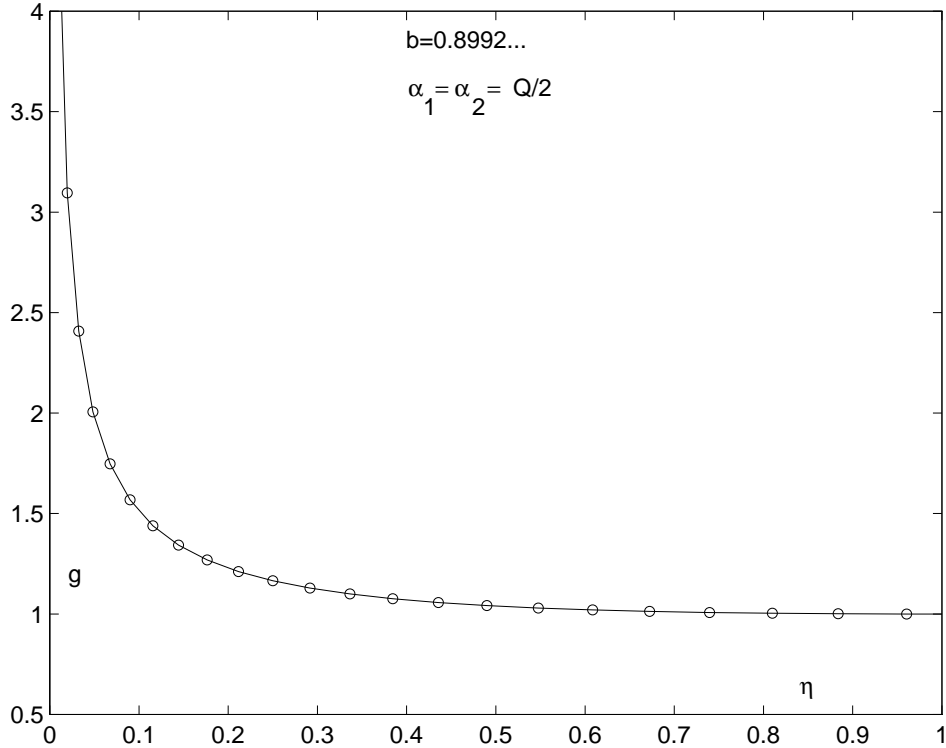}%
\caption{Normalized two-point function $g_{Q/2,Q/2}(\eta)$ evaluated as the
single vacuum block (\ref{F}) (solid line) and as the cross-channel integral
(\ref{Ft}) (circles).}%
\label{fig1}%
\end{center}
\end{figure}

As the first numerical example we take a quite arbitrary value $b^{2}%
=0.8086\ldots$ and two ``puncture'' operators with $\alpha_{1}=\alpha_{2}%
=Q/2$. In this case there are no discrete terms in the expression (\ref{Ft}).
The two-point function $g_{Q/2,Q/2}$ v.s. the invariant ``distance'' $\eta$ is
plotted in fig.\ref{fig1}. Solid line is for eq.(\ref{F}) and circles are
computed as the integral (\ref{Ft}). Few numbers are presented in
table \ref{table1}. The first values $g^{\text{(bound)}}(\eta)$ are evaluated
as the vacuum conformal block (\ref{F}) and $g^{\text{(bulk)}}(\eta)$ stands
for (\ref{Ft}). We quote this table only to illustrate the numerical precision
of our calculations. It should be noted that for $g^{\text{(bulk)}}(\eta)$
only the first 10 digits are correct, the errors being due to the numerical
integration over $P$ in (\ref{Ft}) and evaluation of the special functions
entering the structure constants.

\begin{center}%
\begin{table}[htb] \centering
\begin{tabular}
[c]{|ccc|}\hline
$\eta$ & $g^{\text{(bound)}}(\eta)$ & $g^{\text{(bulk)}}(\eta)$\\\hline
0.10 & 1.511254162734526 & 1.511254162670712\\\hline
0.20 & 1.228318394284875 & 1.228318394218384\\\hline
0.30 & 1.123052815598698 & 1.123052815525115\\\hline
0.40 & 1.069857238682268 & 1.069857238610545\\\hline
0.50 & 1.039506854956745 & 1.039506854882646\\\hline
0.60 & 1.021302866577855 & 1.021302866502048\\\hline
0.70 & 1.010340291843230 & 1.010340291788767\\\hline
0.80 & 1.004036952201786 & 1.004036952278245\\\hline
0.90 & 1.000898855218405 & 1.000898855824359\\\hline
\end{tabular}
\caption{Numerical comparison of the ``boundary'' (\ref{F}) and ``bulk'' (\ref
{Ft})
representations of the normalized two-point function $g_{Q/2,Q/2}(\eta)$ at
$b^2=0.8086\ldots$}\label{table1}%
\end{table}
\end{center}

\begin{figure}
[tbh]
\begin{center}
\includegraphics[
height=4.2186in,
width=5.5365in
]%
{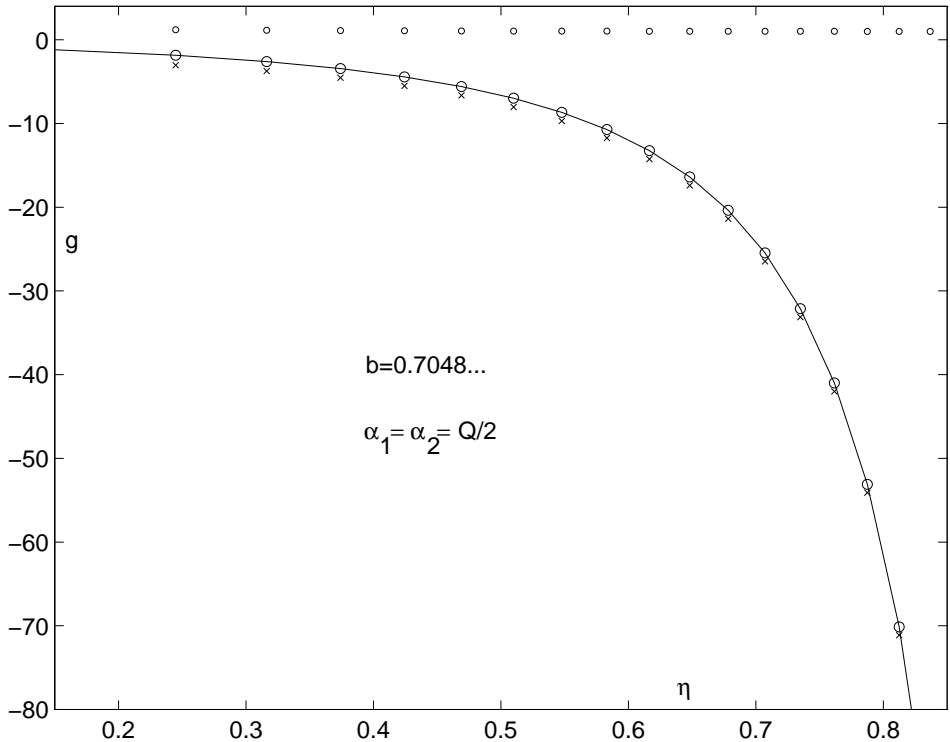}%
\caption{
Boundary (\ref{gFF}) (solid line) and bulk (\ref{Ft}) (circles)
representations of the normalized two-point function $g_{Q/2,Q/2}(\eta)$ are
compared at $b=0.7048\ldots$. Small circles and crosses are respectively the
contributions of the first and second terms in eq.(\ref{gFF}). The two-point
function is almost saturated by the $\psi_{13}$ contribution.
}%
\label{fig3}%
\end{center}
\end{figure}

\textbf{2. Vacuum }$\mathbf{(m,n)=(1,2)}$\textbf{.} In this vacuum two
boundary operators contribute with dimensions $\Delta_{1,1}=0$ and
$\Delta_{1,3}=Q^{2}/4-(b^{-1}+2b)^{2}/4$. Taking again the ``normalized''
correlation function
\begin{equation}
g_{\alpha_{1}\alpha_{2}}(\eta)=\frac{\left\langle V_{\alpha_{1}}(\xi
_{1})V_{\alpha_{2}}(\xi_{2})\right\rangle _{1,2}}{\left\langle V_{\alpha_{1}%
}(\xi_{1})\right\rangle _{1,2}\left\langle V_{\alpha_{2}}(\xi_{2}%
)\right\rangle _{1,2}}\label{gaa12}%
\end{equation}
with respect to this vacuum, we have, instead of eq.(\ref{F})
\begin{align}
\  &  g_{\alpha_{1}\alpha_{2}}(\eta)=(1-\eta)^{2\Delta_{1}}\times\label{gFF}\\
&  \ \ \ \left[  \mathcal{F}\left(
\begin{array}
[c]{cc}%
\alpha_{1} & \alpha_{2}\\
\alpha_{1} & \alpha_{2}%
\end{array}
,iQ/2,1-\eta\right)  +\frac{F_{1,2}(\alpha_{1})F_{1,2}(\alpha_{2})}%
{U_{1,2}(-b/2)F_{1,2}(-b/2)}\mathcal{F}\left(
\begin{array}
[c]{cc}%
\alpha_{1} & \alpha_{2}\\
\alpha_{1} & \alpha_{2}%
\end{array}
,i(b+b^{-1}/2),1-\eta\right)  \right] \nonumber
\end{align}
where
\begin{equation}
F_{1,2}(\alpha)=\frac{B_{1,2}^{(+)}(\alpha)}{U_{1,2}(\alpha)}\label{F12}%
\end{equation}
as given by expression (\ref{Bmn}) with $(m,n)=(1,2).$ The ``cross channel''
representation remains the same as in eq.(\ref{Ft}) with the substitution
$U_{1,1}\rightarrow U_{1,2}$. In fig.\ref{fig3} the numerical values of
(\ref{gFF}) and the cross-channel integral (\ref{Ft}) are compared at
$b=0.7048\ldots$ and again for $\alpha_{1}=\alpha_{2}=Q/2$. Notice that in
this case the two-point function $g_{\alpha_{1}\alpha_{2}}(\eta)$ is an
exponentially growing function of the geodesic distance. This situation is
typical for the ``excited'' vacua $(m,n)\neq(1,1)$ and related to the negative
dimensions (\ref{Dmn}) of the degenerate boundary fields $\psi_{m,n}$ (at real
$b$).%

\section{Discussion}

We have demonstrated that the pseudosphere geometry provides a new physical
picture of 2D quantum gravity. It is different from the compact problems and
in fact much closer to standard physics in ordinary field theory
(peculiarities of a field theory at constant negative curvature are discussed
in \cite{Callan}). However, many conceptual questions related to the suggested
constructions remain open. Let us mention some of them.

\begin{itemize}
\item  Unitary and non-unitary matter fields coupled to Liouville quantum
gravity in this geometry present a separate and rather interesting problem,
particularly in relation to the recent studies of AdS/CFT correspondence
\cite{AdS1}.

\item  Of course the closing of the bootstrap program requires construction of
all bulk-boundary and boundary structure constants for all degenerate boundary
fields associated with different out-vacua $(m,n)$, including the
juxtaposition operators. This difficult technical task remains to be done. For
``ordinary'' boundary conditions in boundary LFT this program has been started
in \cite{Fateev} (see also \cite{Teschner2, Teschner3, Tshit}).

\item  The most intriguing point is the nature of the ``excited'' vacua. As we
have already mentioned, in all such vacua correlation functions typically grow
exponentially with the geodesic distances. This suggests that these states can
be a kind of ``boundary excitations'' of the corresponding boundary conformal
field theory. A meaning of these quantum excitations of the (physically
infinite faraway) absolute remains to be comprehended. Let us mention also
that these growing correlations at large distances are dominated by
non-trivial degenerate boundary operators of negative dimensions. Therefore
the physical ``decay property'' (with which we started our arguments in
sect.2) doesn't hold literally in these excited vacua, being a formal
requirement (\ref{Decay}) for the contribution of the identity operator. This
means that even the logic of the whole development deserves more careful 
examination.
\end{itemize}

\textbf{Acknowledgments.}

Al.Z thanks Department of Physics and Astronomy of Rutgers University, where
this study has been performed. His work was also supported by EU under the
contract ERBFMRX CT 960012. The work of A.Z. is supported by DOE grant
DE-FG02- 96ER10919.

\end{document}